\documentclass[conference]{IEEEtran}
\ifCLASSINFOpdf
   \usepackage[pdftex]{graphicx}
   \graphicspath{{../pdf/}{../jpeg/}}
   \DeclareGraphicsExtensions{.pdf,.jpeg,.png}
\else
\fi
\usepackage{mathtools}
 \usepackage{booktabs}
\usepackage{multirow}
\usepackage{algorithm}
\usepackage{algpseudocode}

\begin{document}
%
\title{A Study of Energy Trading in a Low-Voltage Network: Centralised and Distributed Approaches}



%
\author{\IEEEauthorblockN{Jaysson Guerrero, \textit{Student MIEEE},
Archie Chapman, \textit{MIEEE} and
Gregor Verbi\v{c}, \textit{Senior MIEEE}, 
}
\IEEEauthorblockA{School of Electrical and Information Engineering\\
The University of Sydney, Sydney, New South Wales, Australia\\ Email: \{jaysson.guerrero, archie.chapman, gregor.verbic\}@sydney.edu.au
}}


\maketitle

\begin{abstract}
Over the past years, distributed energy resources (DER) have been the object of many studies, which recognise and establish their emerging role in the future of power systems. However, the implementation of many scenarios and mechanism are still challenging. This paper provides an overview of a local energy market and explores the approaches in which consumers and prosumers take part in this market. Therefore, the purpose of this paper is to review the benefits of local markets for users. This study assesses the performance of distributed and centralised trading mechanisms, comparing scenarios where the objective of the exchange may be based on individual or social welfare. Simulation results show the advantages of local markets and demonstrate the importance of advancing the understanding of local markets.
\end{abstract}

\begin{IEEEkeywords}
	Distributed energy resources, distribution grid, double auction, local market, smart grids.
\end{IEEEkeywords}

%
\IEEEpeerreviewmaketitle

\section*{Nomenclature}
\addcontentsline{toc}{section}{Nomenclature}
\begin{IEEEdescription}[\IEEEusemathlabelsep\IEEEsetlabelwidth{$V_1,V_2,1$}]
	\item[$k$] Time-slot 
\item[$\mathcal{K}$] Set of time-slots $k$             
\item[$x^\mathrm{j,PV}_{k}$] PV generation at time ${k}$, user ${j}$  
\item[$x^\mathrm{j,Batt}_{k}$] Energy stored at time ${k}$, user ${j}$                      
\item[$\mathcal{J}$] Set of users $j$ 
\item[$\mathcal{B}$] Set of buyers $b$
\item[$\mathcal{S}$] Set of sellers $s$
\item[$\mathcal{P}$] Set of prices $p$
\item[${p}_{b}$] Offer price of buyer $b$ 
\item[${p}_{s}$] Bid price of seller $s$
\item[$\mathcal{Q}$] Set of quantities of energy $q$ 
\item[${q}_{b}$] Quantity of energy to purchase by buyer $b$ 
\item[${q}_{s}$] Quantity of energy to supply by seller $s$
\item[${t}_{d}$] Trading time
\item[$\mathcal{ID}$] Set of identification index $id$
\item[$\mathcal{OB}$] Set of all orders in the order book
\item[$Mo^{\delta}$] Market order with index $\delta$ 
\item[$p_{t}$] Transaction price	
\item[$q_{t}$] Quantity of energy to exchange
\item[$Q_{T}$] Total of energy traded
\item[$U$] Utility function	
\item[$C$] Cost function
\item[$L_{min}$] Minimum value of trading prices 
\item[$L_{max}$] Maximum value of trading prices
\end{IEEEdescription}

\section{Introduction}
Increasing penetration of renewable electricity generation and energy storage technology characterise the future of electrical power systems. All these developments along with an advanced network communication constitute one part of the smart grid vision. Undoubtedly, these technologies will bring more active participation of end users. As a consequence, the network will confront changes in its structure as well as in the business model. 

Australia is one of the leaders in PV installation around the world. Around 900,000 rooftop PV systems were installed in Australia between 2010 and 2012 \cite{mountain_chapter_2014}. The expansion of PV is increasing as a result of some factors such as the rising rates of electricity, subsidies and advances in technologies, which are bringing more profitable solutions for users. However, incentives remain under the expectations of customers. For example, one of the most common production subsidies for energy users is feed-in-tariff, in which households will receive payments for the power exported to the grid. Nevertheless, this subside may not cover the revenue desired to recover the initial investment and the cost associated with energy generation. Users with PV systems and battery storages may take advantage of their surplus of energy, optimising their energy consumption. Hence, users in a smart grid would be willing to seek more profitable alternatives to the current business model and to participate in a more efficient model.

A clear example of innovation in this area is the pilot project named \textit{deX} (decentralised energy exchange), which is funded by the Australian Renewable Energy Agency (ARENA) and led by GreenSync \cite{deX}. The aim of that project is an online exchange platform for buying and selling grid-services such as power from distributed energy resources. This scenario brings benefits for consumers through reduced electricity network bills. Furthermore, technological developments have brought more tools to facilitate the implementation of these models. For instance, recently some studies have considered using blockchain technology and smart contracts in electricity markets \cite{mihaylov_nrgcoin:_2014}, \cite{munsing_blockchains_2017}. A platform based on blockchain may be used to enhance the security of transactions, through a virtual currency, in a local energy market. Despite new enabling technologies, their complete deployment is not clear yet. Therefore, many emerging scenarios and uncertainties remain unsolved. 

In the context of a local low-voltage network with a small group of electricity users, the vision of a local trading market could be established as one alternative to the current business models. In \cite{bayram_survey_2014}, an overview of distributed energy trading in smart grid is presented. That survey explores the existing literature of trading algorithms involved in market frameworks such as distributed, centralised and simulation-based solutions. Likewise, the application and features of a local electricity market have been identified in \cite{ampatzis_local_2014}. In fact, these studies have shown that consumers and prosumers may perform a local market in order to obtain profitable benefits which depend on factors such as load profiles, energy surplus and fair prices. Within this context, previous research has explored the opportunity of trading in micro-grid networks. For example, in \cite{matamoros_microgrids_2012}, the case of energy trading among isolated micro-grids have been addressed. To this end, the authors consider centralised as well as distributed approaches as minimisation cost problem. Similarly, in \cite{cui_electricity_2014-1}, welfare maximisation problem is described to deal with energy trading among micro-grids. While these studies focussed on interaction between networks, a direct participation of low-voltage network users in a local market may also be established.

In this paper, we consider that the units to be traded are the result of a surplus of energy generated and/or stored. As discussed in \cite{ilic_energy_2012}, \cite{endo_distributed_2017}, numerous approaches for energy trading have been introduced considering different types for price discovery process. Independently whether the structure in the market is centralised or not, it has been demonstrated that a scenario of prosumers with a surplus of energy and consumers in a local market is viable and reaches high levels of efficiency. In the case of distributed context, via a double auction, the prices may be determined as a consequence of offers submitted continuously by agents. Through this mechanism, all agents may achieve some benefits even though they do not have a specific bidding strategy. Similarly, other works have shown the savings achieved by energy trading with a game theoretic approaches \cite{wang_game-theoretic_2014}, \cite{yaagoubi_energy_2015}. These studies encourage the use of bidding mechanisms for this context.

Given these insights, the overall contribution of this paper is an analysis of energy trading in a local market, where prosumers and consumers are able to offer and to purchase energy. This paper seeks partially to bridge energy management system and bidding mechanism in a local market. Firstly, users minimize the cost of self-consumption, and subsequently they identified their amounts and times to trade. Specifically, we considered centralised and distributed approaches as electricity market mechanism. The last one is based on the continuous double auction, which has been defined as high efficient method by previous studies \cite{nicolaisen_market_2001}. Moreover, we consider centralised and distributed approaches in order to compare the benefits of each scheme. In doing so, we illustrate the prices and the amounts of energy exchanged among agents in the market.

This paper progresses as follows: The next section of the paper states the system model. This is followed by the description of the implementation in Section III. Then, Section IV presents simulations results and the discussion. Finally, Section V concludes.

\section{System Model}
\begin{figure}[h]
	\centering
	\includegraphics[width=2in]{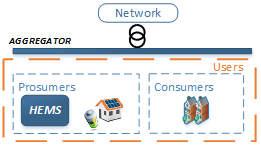}
	\caption{Users in a low-voltage network}
	\label{scenario}
\end{figure}

Our study is focused on a low-voltage network with \textit{distributed energy resources} (DER), as shown in Fig \ref{scenario}. While one part of the households (prosumers) have PV systems, battery storage and \textit{home energy management systems} (HEMS); the other group is constituted by traditional customers willing to pay rates defined by the grid operator.

There are three components in our model. The first one is the local power network, the second one is the customers and the last one is the market for energy trading.

\subsection{Local Power Network Model}
We consider a smart grid system for energy trading at local level. In order to minimize the energy costs associated to the use of energy, prosumers in the network have PV systems, battery storage and HEMS. Let $\mathcal{J}$ denotes the set of all $j$ users in the local grid. The time is divided into time slots $k \in \left \{  1,2,...,\mathcal{K}\right \}$, where $\mathcal{K}$ is the total number of time slots. We define $x_{k}^{j}$ as the amount of energy used by the user $j \in {\mathcal{J}}$ in time slot $k$. There are two categories of users in power systems, consumers and prosumers. The model of the users is based on CREST model \cite{mckenna_high-resolution_2016}, , which is a high-resolution stochastic model of electricity demand. This model simulates electrical demand and generation due to appliances, lighting and photovoltaics systems. The first objective of the prosumers in $\mathcal{J}$ is to optimise their self-consumption considering their demand and the energy generated through the PV system $x_{k}^{j,PV}$, and the energy stored in the battery $x_{k}^{j,Batt}$. Therefore, the optimisation problem of a HEMS is given by:

\begin{equation}\label{HEMS}
	\begin{aligned}
	& \underset{X}{\text{min}}
	& & \sum_{k=1}^\mathcal{K}{(s_{k}^{+}x_{k}^{+} - s_{k}^{-}x_{k}^{-}}) \\
	& \text{s.t.}
	& & \text{satisfies storage device, comfort, power flow} \\
	&&& \text{and energy balance constraints,} \\
	&&& \forall{k} \in{\left \{1... \mathcal{K}\right \}}\\
	\end{aligned}
\end{equation}

\noindent where, $x_{k}^{+}$ and $x_{k}^{-}$ are the amount of energy flowing from the grid and to the grid respectively. State variables in the model are $s_{k}^{+} $ and $s_{k}^{-}$. The former is associated with the price of energy in time slot $k$, and the latter with the incentive received for the contribution to the grid. In other words, $s_{k}^{+} $ and $s_{k}^{-}$ may be related to rates (e.g. flat, time-of-use) or incentives (e.g. feed-in-tariff). The outcome of the previous process provides \textit{net load profiles} for users with HEMS. Given the prosumers have an excess of energy after their self-optimisation, a local market for energy trading may be established.

\subsection{Energy Trading Model}

 The effectiveness and performance of the market depends on the mechanisms implemented. In our study, we have considered distributed and centralised schemes. 

\subsubsection{Distributed Market}

\begin{figure}[h]
	\centering
	\includegraphics[width=1.5in]{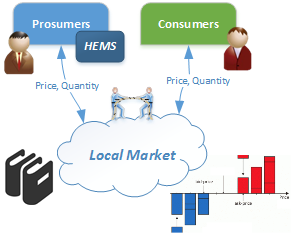}
	\caption{Prosumers and Consumers submit continually their offers and bids.}
	\label{market}
\end{figure}

The operation of the market in this case involves only two parties interested in the trading. Hence it is peer-to-peer (P2P) and bilateral contract between agents (buyers and sellers). In order to achieve an individual welfare, the agents submit offers/bids based on their preferences and costs. The local market is based on a \textit{continuous doubled auction} (CDA), where there are a set of buyers, $b\in{\mathcal{B}}$, and sellers, $s\in{\mathcal{S}}$, willing to participate continuously in the market considering their trading prices ($p_{b}$, $p_{s}$) and their amount of energy to purchase or supply ($q_{b}$, $q_{s}$) (See Fig \ref{market}). Previous studies have shown that market efficiency may be directly attributed to continuous double auctions \cite{nicolaisen_market_2001}. This auction mechanism is widely used in stock markets around the world. Some examples are the NASDAQ, the New York Stock Exchange and online markets such as the auctions conducted by eBay. In a CDA, the agents offer or bid during a trading time $t_{d}$ and their offers and bids are registered in a \textit{order book} $\mathcal{OB}(id,p,q,t)$, where each order has an index $id \in{\mathcal{ID}}$, price $p \in{\mathcal{P}}$, quantity $q \in{\mathcal{Q}}$ and the time when the order was received $t\in{\mathcal{T}}$. During the trading time, the process of arrivals in the order book follow a Poisson process with a mean $\lambda$. Additionally, this model is a multi-unit market where units exchanged symbolize the flowing of power between agents, which is their main motivation to participate in the market.

 Let define $Mo^{\delta} \in{\mathcal{OB}}$ as the market order with index ${\delta \in{\Delta}}$. An order is a market order if this one had a match during the trading time $t_{d}$. Once the market is closed, the outcome of the trading is a set of market orders $Mo^{\delta}$ with a transaction price $p_{t}^{\delta}$ and quantity $q_{t}^{\delta}$ .For the matching process, there are two fundamental properties in offers/bids to be considered. The first one is the price, and the second one is the time in which the offer/bid was recorded in the order book. Hence, the best offer (buyers) is the earliest offer with the highest price. Likewise, in the case of bids (sellers), the best one is the earliest bid with the lowest price. To determine whether a transaction is completed or not, the best bid and the best offer are compared. If $p_{b} \geq p_{s}$, the orders are matching and the agents will exchange energy. Otherwise, it will remain in the order book. If a new offer/bid is not better than the best one, it will be aggregated to the order book regarding its arrival time and price. This process is executed several times in the order book during the trading period. After the matching process, an order can have covered their request partially . If this is the case, it will remain at the top of the order book waiting for a new order.
 
Once the trading time has elapsed, the total of energy $Q_{T}$ is given by:

\begin{equation}
Q_{T} = \sum_{\delta\in{\Delta}}{q_t^{\delta}}
\end{equation}

Conventionally, the participants of markets, buyers and sellers, define their offers and bids based on their preferences and costs associated. Since our interest is to assess the benefits of a local market, in our study the agents are \textit{zero intelligence plus} (ZIP) traders. In \cite{gode_allocative_1993}, Gode and Sunder designed zero intelligence traders. Buyers and sellers submit randomly their offers or bids depending on their constraints ($L_{max}$ and $L_{min}$ are the maximum and minimum price respectively). In order to improve the performance of zero intelligence traders, Cliff and Bruten \cite{ZIP} developed ZIP traders. Agents have a profit margin which determines the difference between their limit prices and their offers or bids. Under this strategy, traders adapt and update their margins base on the matching of previous orders. The algorithm \ref{zi} shows an overview of the order book process with ZIP traders.  

\begin{algorithm}
	\caption{Algorithm for the ZIP traders}
	\label{zi}
	\begin{algorithmic}[1]
		\Procedure{Order Book}{$\mathcal{OB}$}
		\State initialization;
		\While {market is open}
		\State randomly select a new trader
		\State new order by ZIP-trader
		\If{buyer} 
		\State	new $\mathcal{OB}(id_{b},p_{b},q_{b},t)$ 
		\Else 
		\State	new $\mathcal{OB}(id_{s},p_{s},q_{s},t)$ 
		\EndIf
		\State 	allocation of new order in $\mathcal{OB}$
		\State  evaluate matching process
		\State \Comment{Update values of profit margins ---------- Buyers}
		\If{the last order was matched at price $q_{t}$} 
		\State	all buyers for which $p_{b}\geq q_{t}$, raise his margin;  
		\If {the last trader was a seller}
		\State any active buyer for which $p_{b}\leq q_{t}$, 
		\State lower his margin;
		\EndIf
		\Else 
		\If {the last trader was a buyer}
		\State any active buyer for which $p_{b}\leq q_{t}$, 
		\State lower his margin;
		\EndIf			
		\EndIf
		\State \Comment{Update values of profit margins ---------- Sellers} 
		\If{the last order was matched at price $q_{t}$} 
		\State	all sellers for which $p_{s}\leq q_{t}$, raise his margin;  
		\If {the last trader was a buyer}
		\State any active seller for which $p_{b}\geq q_{t}$, 
		\State lower his margin;
		\EndIf
		\Else 
		\If {the last trader was a seller}
		\State any active seller for which $p_{b}\geq q_{t}$, 
		\State lower his margin;
		\EndIf			
		\EndIf
		
		\EndWhile		
		\EndProcedure
	\end{algorithmic}
\end{algorithm}

\subsubsection{Centralised Market}

In this structure, the optimal dispatch is decided based on all information of consumers and prosumers. In order maximize the global welfare, the agents will develop an energy allocation algorithm to identify the market equilibrium in each trading period. Commonly, the social welfare is formulated through the utility and cost functions. Each consumer has a utility function $U(q_{b})$ that models the level of satisfaction due to purchase energy. Likewise, each prosumer has a cost function $C(q_{s})$ that represents the costs associated to the amount generated. Regarding the social welfare problem must ensure power balance, the maximisation of social welfare takes the form:

\begin{equation}
\begin{aligned}
& \underset{q_{b}, q_{s}}{\text{max}}
& & \sum_{t \in \mathcal{T}}{U(q_{b}^{t}) - C(q_{s}^{t})} \\
& \text{s.t.}
& & \text{units constraints and load balance} \\
&&& \text{constraints} \\
\end{aligned}
\end{equation}

\noindent While the dispatch is decided through optimal allocation, the methodology for pricing  depends on the mechanism to define one \textit{market clearing price} (MCP). The value of this variable represents the price of each unit to be exchanged. Generally, MCP is equal to the equilibrium price, the intersection of supply and demand curves. As a result of this process, all bids with $p_{s} \leq MCP$, as well as all offers with $p_{b} \geq MCP$ are accepted. Consequently, consumers and prosumers are informed of the amount to be traded.

Although this mechanism leads to a balance in the market between agents, different methods have been developed to bring other properties to the markets. In particular, Vickery-Clarke-Groves (VCG) Mechanism is incentive compatible for optimising the social welfare through efficient allocation where each agent pays an amount equal to the social cost/damage that he causes the other players \cite{tardos_introduction_2007}. Hence, truthfulness is a dominant strategy in this mechanism. In order to compare two approaches in this centralised market, we will evaluate these mechanisms in the formulated scenario.


As this is a preliminary study, we do not yet consider network losses or constraints, in effect assuming a copper-plate network model. A full network representation will be integrated into the trading mechanism as part of future work.

\section{Implementation}

\begin{figure}[h]
	\centering
	\includegraphics[width=2.7in]{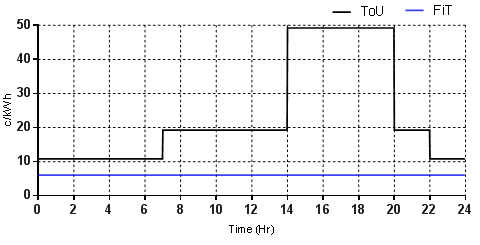}
	\caption{Time-of-use tariff (ToU) and feed-in-tariff (FiT)}
	\label{tariffs}
\end{figure}

The proposed market consists of a group of dwellings, comprising a mix of consumers and prosumers, a market structure determining prices and matching of trades, a copper-plate network facilitating transport of energy, and a communications network enabling the flow of market-related information. Demand profiles, with $k=1$ minute resolution, are based on CREST Demand Model \cite{mckenna_high-resolution_2016}. In Fig \ref{tariffs} are depicted the time-of-use tariff (ToU) and feed-in-tariff (FiT) used in our model. Since ZIP traders improved their performance when the maximum and minimum constraints are defined, we use the values of tariff through the day to define $L_{max}$ and $L_{min}$. Hence, the former depends on the ToU tariff and the latter on FiT. These definitions are consistent in the sense that no buyer would pay more than the tariff of the retailer (ToU) and no sellers would sell their units cheaper than the incentive (FiT) that they would expect to receive. In summary, the process of our model is:

\begin{description}
	\item[$\bullet$] Prosumers run HEMS to minimize their cost based on the optimisation problem (\refeq{HEMS}) formulated previously. To solve the problem, we used Mixed-integer Linear Program (MILP).
	\item[$\bullet$] Prosumers state the time-slots when they have extra energy to trade.
	\item[$\bullet$] The input data for the market include load profiles of prosumers and consumers, and the value of the tariff.
	\item[$\bullet$] Local market is performed and Order Book starts to receive orders during trading periods.
	\item[$\bullet$] Agents accept the amount of units to be exchanged and their prices.
	
\end{description}

Additionally, two scenarios were evaluated. In the first one, the energy to trade is a consequence of extra energy generated by PV systems. In the second, prosumers are willing to trade their surplus of energy generated as well as stored in their batteries. More specifications and results are explained in the next section. 

\section{Simulation Results and Discussion}

For simulating the proposed system, we regarded load profiles from CREST Model of 100 dwellings, of which 37 are prosumers and 63 are consumers. Preceded by HEMS process, prosumers identified their surplus, and therefore quantities to trade in the market. 

\subsection{Scenario 1}

In this case, the best time for participating in the market is around middle day as a consequence of extra units generated by PV systems. Moreover, users with HEMS meet their demand and store energy in their battery to use at peak time prices. For this reason, in our case-study, the most productive periods to trade were established between 8 am and 3 pm. As it was mentioned before, the market was performed considering three mechanisms: Centralised with equilibrium price, centralised with VCG mechanism and distributed P2P market. To compare the prices of transactions in each mechanism, we calculated the average transaction price $\left<{T_{p}}\right>$ during the day. 

\begin{figure}[h]
	\centering
	\includegraphics[width=3.5in]{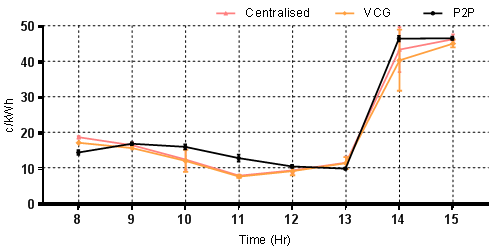}
	\caption{Average transaction prices $\left<{T_{p}}\right>$ in scenario 1}
	\label{AVP}
\end{figure}

Fig \ref{AVP} shows the average transaction price each hour. The trend of all prices is to remain in the range of $L_{min}$ and $L_{max}$ (i.e. values of ToU and FiT). Hence, both buyers and sellers obtain a benefit from the local market. In the context of P2P case, there are no large fluctuations during each trading period. This is due to the strategy used by ZIP traders, agents learned during trading and modified their margins to participate in the market. The transactions prices converge rapidly result in no significant variations. The number of traders and the units to trade are different each trading period. Therefore, the transaction price does not necessarily have to converge to the same value. There is a peak at 2 pm because of the change in the tariff at that time.  Additionally, we can conclude that VCG mechanism brings slightly lower prices in a centralised case. However, the cause of this is potentially associated with one feature of VCG mechanism which is budget deficit. In this case, sellers accepted prices in benefit of buyers. Those prices tend to be less than the equilibrium price. 

\begin{figure}[h]
	\centering
	\includegraphics[width=3.5in]{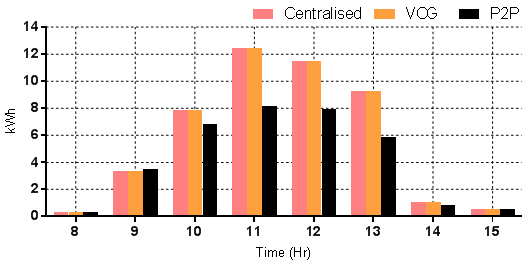}
	\caption{Total energy traded $Q_{T}$ in scenario 1}
	\label{TE}
\end{figure}

In the Fig \ref{TE}, the total energy traded $Q_{T}$ each hour is presented. In the first hours, the amount traded is small and increases gradually to reach a peak around middle day. Finally, it decreases progressively over time. Furthermore, the quantity exchanged with P2P mechanism is slightly less during the whole day. This is caused by factors associated with the auction process and agents strategies such as arrival orders time, margin prices and the evolution of the learning process by ZIP traders. In the case of centralised options, the units are the same because they use the same method for the allocation.

The Table \ref{t1} shows savings and profits (in dollars) that consumers and prosumers would have potentially with their participation in the local market during one day. Additionally, the table compares values to show the profitable of each mechanism. Savings indicates the money that buyers are economising due to buying in the local market instead of the grid operator. Similarly, profit represents the extra money that sellers would earn.

\begin{table}[h]
	\centering
	\caption{Scenario 1. Simulation results for different mechanisms in the local market. These values represent the revenues for agents}
	\label{t1}
	\begin{tabular}{c|crcccc}
		\hline
		\multirow{2}{*}{\textbf{\begin{tabular}[c]{@{}c@{}}Time \\ (Hr)\end{tabular}}} & \multicolumn{2}{c|}{\textbf{Centralised}} & \multicolumn{2}{c|}{\textbf{VCG}} & \multicolumn{2}{c}{\textbf{P2P}} \\ \cline{2-7} 
		& \textbf{Savings} & \textbf{Profit} & \textbf{Savings} & \textbf{Profit} & \textbf{Savings} & \textbf{Profit} \\ \hline
8 & 0.01 & 0.04 & 0.01 & 0.04 & 0.01 & 0.01 \\
9 & 0.10 & 0.34 & 0.12 & 0.32 & 0.06 & 0.29 \\
10 & 0.61 & 0.42 & 0.65 & 0.39 & 0.14 & 0.64 \\
11 & 1.42 & 0.21 & 1.46 & 0.18 & 0.45 & 0.67 \\
12 & 1.15 & 0.35 & 1.18 & 0.33 & 0.57 & 0.39 \\
13 & 0.74 & 0.47 & 0.76 & 0.46 & 0.64 & 0.30 \\
14 & 0.11 & 0.34 & 0.17 & 0.30 & 0.10 & 0.36 \\
15 & 0.01 & 0.20 & 0.02 & 0.20 & 0.02 & 0.27 \\ \hline
	\textbf{Total} & \textbf{4.19} & \textbf{2.39} & \textbf{4.37} & \textbf{2.21} & \textbf{1.98} & \textbf{2.93} \\ \hline
	\end{tabular}
\end{table}

From the Table \ref{t1}, we can see that there are benefits for all agents in the market regardless of the mechanism used. For the buyers, centralised mechanisms were more profitable, particularly with VCG mechanism. In the case of the sellers, they achieved more profit from a distributed mechanism. Even although with the centralised mechanism the amount of energy was greater than in the distributed mechanism, the payoff may be higher because of the transaction price.

\subsection{Scenario 2}

\begin{figure}[h!]
	\centering
	\includegraphics[width=3.5in]{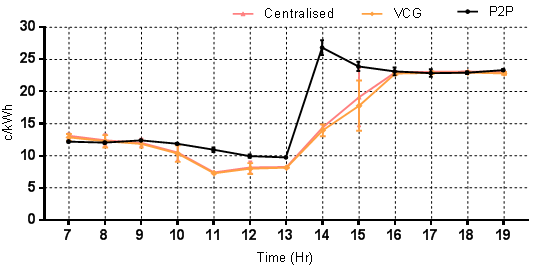}
	\caption{Average transaction prices $\left<{T_{p}}\right>$ in scenario 2}
	\label{AVP2}
\end{figure}

\begin{figure}[h!]
	\centering
	\includegraphics[width=3.5in]{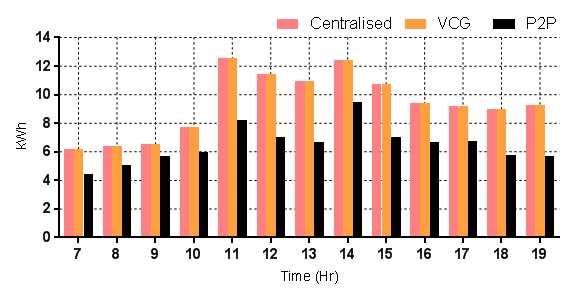}
	\caption{Total energy traded $Q_{T}$ in scenario 2}
	\label{TE2}
\end{figure}

In this case, prosumers are willing to keep some energy in the battery to trade instead of using the whole surplus only for their self-consumption. Consequently, there will be others trading times throughout the day. Results of this scenario are shown in figures \ref{AVP2} and \ref{TE2}.

An increasing of hours for trading is evident. Similar to the previous scenario, there are no significant changes during each trading periods, and the prices respond to each mechanism and agents strategies. Likewise, the change in the value of the tariff causes high variation in prices around 2 pm. The total of energy traded in this scenario is substantially more. Meanwhile, in the first scenario there was no energy to trade after 3 pm; in this case, the time to trade was extended until 7 pm. Therefore, sellers have the opportunity to trade during more time and buyers may avoid peak prices from the grid operator. After the peak time, sellers may start to charge their batteries again at better prices (shoulder and off-peak periods). As shown in Table \ref{t2}, the more profitable method, for both sellers and buyers, are the centralised mechanisms. However, there were some time periods when the P2P mechanism was better for sellers (from 11 am to 3 pm).


\begin{table}[h]
	\centering
	\caption{Scenario 2. Simulation results for different mechanisms in the local market. These values represent the revenues for agents}
	\label{t2}
\begin{tabular}{c|cccccc}
	\hline
	\multirow{2}{*}{\textbf{\begin{tabular}[c]{@{}c@{}}Time \\ (Hr)\end{tabular}}} & \multicolumn{2}{c|}{\textbf{Centralised}} & \multicolumn{2}{c|}{\textbf{VCG}} & \multicolumn{2}{c}{\textbf{P2P}} \\ \cline{2-7} 
	& \textbf{Savings} & \textbf{Profit} & \textbf{Savings} & \textbf{Profit} & \textbf{Savings} & \textbf{Profit} \\ \hline
	7 & 0.38 & 0.43 & 0.40 & 0.42 & 0.24 & 0.20 \\
	8 & 0.44 & 0.40 & 0.45 & 0.39 & 0.34 & 0.29 \\
	9 & 0.47 & 0.39 & 0.48 & 0.38 & 0.39 & 0.33 \\
	10 & 0.69 & 0.32 & 0.71 & 0.31 & 0.39 & 0.32 \\
	11 & 1.49 & 0.16 & 1.50 & 0.15 & 0.69 & 0.44 \\
	12 & 1.27 & 0.23 & 1.29 & 0.21 & 0.68 & 0.30 \\
	13 & 1.20 & 0.24 & 1.21 & 0.23 & 0.61 & 0.23 \\
	14 & 4.32 & 1.04 & 4.38 & 0.98 & 1.78 & 1.63 \\
	15 & 3.28 & 1.36 & 3.42 & 1.22 & 1.94 & 1.42 \\
	16 & 2.47 & 1.60 & 2.49 & 1.58 & 1.67 & 1.12 \\
	17 & 2.40 & 1.56 & 2.41 & 1.54 & 1.68 & 1.04 \\
	18 & 2.35 & 1.53 & 2.36 & 1.52 & 1.58 & 1.01 \\
	19 & 2.44 & 1.58 & 2.46 & 1.56 & 1.59 & 1.04 \\ \hline
	\textbf{Total} & \textbf{23.22} & \textbf{10.83} & \textbf{23.57} & \textbf{10.49} & \textbf{13.57} & \textbf{9.40}
\end{tabular}
\end{table}

Both scenarios have shown profitable results for traders. Regardless of the mechanism used, consumers and prosumers have a great opportunity to achieve revenues if they perform and participate in a local market.

\section{Conclusion}
In this paper, we have assessed the performance and profitable of a local market. This has been formulated considering centralised and distributed mechanism in a continues double auction. Units traded represent the energy surplus of a group of users with PV systems, battery storage and HEMS. Our simulations and results have shown the benefits that the local market will bring to their participants independently of the mechanism implemented. For future work, it is of interest to extend the study of strategies of agents with a more extensive analysis including penalties and technical constraints in the network.






\bibliographystyle{IEEEtran}
\bibliography{test}
%
	
	

\end{document}